\documentclass[twocolumn,aps,pra,superscriptaddress]{revtex4-1}
\usepackage{hyperref}
\usepackage{verbatim}
\usepackage{amsmath}
\usepackage{latexsym}
\usepackage{revsymb}
\usepackage{yfonts}
\usepackage{ifthen}

\usepackage{natbib}
\usepackage{amsfonts}
\usepackage{amsmath}
\usepackage{amssymb}
\usepackage{amsthm}
\usepackage{graphicx}
\usepackage{bm}
\usepackage{bbm}
\usepackage{epsfig,color,amssymb}
\usepackage{subfigure}
\usepackage{amsfonts}
\usepackage{amscd}
\usepackage{amsmath}
\usepackage{multirow}
\usepackage{chemarrow}
\usepackage{dcolumn}
\usepackage{bm}
\usepackage{graphicx}
\usepackage{enumerate}
\usepackage{epsfig}
\usepackage{subfigure}
\usepackage{xcolor}
\usepackage{multirow}
\usepackage{braket}
\usepackage{comment}
\usepackage{enumitem}
\usepackage{amsthm}
\renewcommand{\emph}[1]{\textit{#1}}

\begin{document}

\title{Differential phase shift quantum secret sharing using twin field}

\author{Jie Gu}
\author{Xiao-Yu Cao}
\author{Hua-Lei Yin}\email{hlyin@nju.edu.cn}
\affiliation{National Laboratory of Solid State Microstructures, School of Physics and Collaborative Innovation Center of Advanced Microstructures, Nanjing University, Nanjing 210093, China}
\author{Zeng-Bing Chen}\email{zbchen@nju.edu.cn}
\affiliation{National Laboratory of Solid State Microstructures, School of Physics and Collaborative Innovation Center of Advanced Microstructures, Nanjing University, Nanjing 210093, China}



\begin{abstract}
Quantum secret sharing (QSS) is essential for multiparty quantum communication, which is one of cornerstones in the future quantum internet. However, a linear rate-distance limitation severely constrains the secure key rate and transmission distance of QSS. Here, we present a practical QSS protocol among three participants based on the differential phase shift scheme and twin field ideas for the solution of high-efficiency multiparty communication task. In contrast to formerly proposed differential phase shift QSS protocol, our protocol can break the linear PLOB bound, theoretically improving the secret key rate by three orders of magnitude in a 300-km-long fiber. Furthermore, the new protocol is secure against Trojan horse attacks that cannot be resisted by previous differential phase shift QSS.
\end{abstract}
\maketitle

\section{Introduction}
Secret sharing is one of basic communication missions in classical cryptography. It aims to split a message into several parts in such a way that no unauthorized subset is sufficient to reconstruct the original message~\cite{shamir1979share}. Since it feathers in secure collaborative activities, secret sharing constitutes a major cryptographic mission for multiparty communication, including controlling missiles\cite{simmons1989prepositioned}.

All schemes of classical secret sharing can only be proved secure based on mathematical complexity. Thus, with the emergence of quantum computing algorithm\cite{shor1994algorithms}, secret sharing is also forced to step into the information-theoretically secure era with laws of quantum mechanics, which is called quantum secret sharing (QSS). The first quantum secret sharing protocol was proposed in 1999 using three-photon entangled Greenberger-Horne-Zeilinger (GHZ) state for three participants~\cite{hillery1999quantum}. Based on multipartite quantum entanglement, improved quantum secret sharing protocols~\cite{xiao2004efficient,markham2008graph,Kogias:2017:Unconditional,9006878,kang2020quantum,moreno2020device} and experimental demonstrations\cite{chen2005experimental,gaertner2007experimental,bell2014experimental,Zhou:2018:Quantum} have been presented in the past 20 years. Nevertheless, difficulties in preparing and transmitting GHZ states constrains the secret key rate and stability of QSS systems, making it far from practical implementation. Therefore, several protocols in the prepare-and-measure scenario were proposed to circumvent preparations of multipartite entangled states, such as protocols using post-selected multipartite entanglement state~\cite{fu2015long}, Bell states~\cite{karlsson1999quantum,SUN20093647}, continuous variable~\cite{Warren:2019:Quantum,wu2020passive}, single-qubit scheme~\cite{schmid2005experimental,bogdanski2008experimental,pinnell2020experimental,de2020experimental}, d-level scheme~\cite{karimipour2015quantum,tavakoli2015secret} and differential phase shift scheme~\cite{inoue2008differential}. Unfortunately, a majority of prepare-and-measure protocols~\cite{schmid2005experimental,bogdanski2008experimental,karimipour2015quantum,tavakoli2015secret,inoue2008differential,pinnell2020experimental,de2020experimental} is vulnerable to the Trojan horse attacks\cite{gisin2006trojan,jain2014trojan}. Furthermore, in particular, all proposed QSS protocols will confront the linear rate-distance limitation~\cite{takeoka2014fundamental,pirandola2017fundamental,das2019universal}. This limitation constricts the key rate and transmission distance of QSS.

Consequently, although QSS has received a lot of attention and achieved a lot of research results, a practical and high efficiency protocol is still missing. In this paper, we present a QSS protocol inspired by the idea of twin-field~\cite{lucamarini2018overcoming,curty2019simple,yin2019finite} and differential phase shift~\cite{inoue2003differential} quantum key distribution. Secure against individual attacks, our protocol will break the linear Pirandola-Laurenza-Ottaviani-Banchi (PLOB) bound~\cite{pirandola2017fundamental}. Besides, in contrast to the previous differential phase shift QSS~\cite{inoue2008differential}, our protocol realizes that the final key rate can be increased by three orders of magnitude in a 300-km-long fiber and secure against Trojan horse attacks. Additionally, in our protocol, senders only utilize weak coherent laser sources to generate signal pulses. Simultaneously, the phase stabilization method will be employed, which has been widely developed in twin-field quantum key distribution~\cite{liu2019experimental,minder2019experimental,fang2020implementation}. With simple apparatus requirements, our protocol will lower the threshold of experimental and practical implementations.

\section{Protocol Description}

\begin{figure}[t!]
  \centering
  \includegraphics[width=86mm]{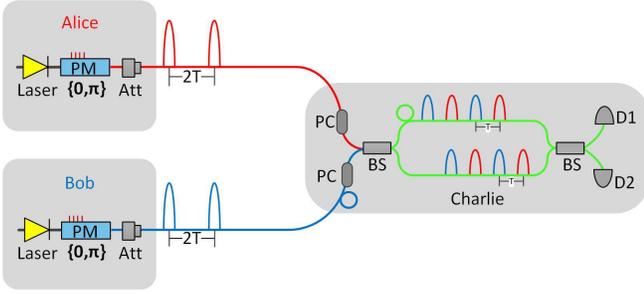}
  \caption{\textbf{Configuration of our quantum secret sharing protocol.}
  Weak coherent pulse sources (Laser);  phase modulator (PM); signal attenuator (Att); polarization control (PC); beam splitter (BS); detector (D1, D2); In Charlie's measurement area, two pulse trains are first been polarization-modulated by polarization controls to correct their polarization for interference.}
  \label{setup}
\end{figure}

To introduce our protocol, we plot in Fig. \ref{setup} the configuration of our quantum secret sharing protocol under experimental setup. In Fig. \ref{setup}, Charlie, the dealer~\cite{gaertner2007experimental}, holds a full key for ciphering while Alice and Bob each hold a partial key for deciphering. In grey areas locate devices held by three participants which are all secure against any eavesdropping. Our protocol employs weak coherent states instead of single-photon states. It features a simple setup where senders basically require weak coherent light sources to generate pulse trains. Furthermore, the dealer utilizes an asymmetric Mach-Zehnder interferometer to measure, with the phase stabilization method~\cite{liu2019experimental,minder2019experimental,fang2020implementation}. The explicit description of our protocol is presented step by step as follows: we denote the sequence of time slots after combination of pulses in two paths as $j$, for $j \in \{1,2,3,...,2N\}$ where $N$ is the total number of pulses sent by Alice (or Bob) and We set that $k$ is an integer ranging from $1$ to $N$.

~\noindent{\it{1.~Preparation.}} Alice (Bob) sends a weak coherent pulse train to Charlie where each pulse is randomly phase-modulated  by $\{0, \pi\}$, respectively. The period of the pulse train is $2T$, as shown in Fig.\ref{setup} and the intensity of each pulse is $\mu$, which also represents the average photon number of each number. $\mu$ is set less than one photon per pulse. Alice (Bob) records their logic bits of each time slot as 0 (1) when her (his) modulated phase is $0~(\pi)$.

~\noindent{\it{2.~Measurement and reconciliation.}} Before the former beam splitter, the pulse train that Bob sends will be postponed for the time $T$. Then Charlie measures the phase difference of adjacent pulses with an $T$ time  delayed  Mach-Zehnder interferometer. He records which detector clicks and the corresponding detection time slot. His logic bits will be 0 (1) when D1 (D2) clicks. For double clicks, Charlie randomly chooses a logic bit out of 0 or 1. Then, Charlie discloses the photon detection time. With time information, Alice, Bob and Charlie form their own raw key for the next procedure.

~\noindent{\it{3.~Parameter estimation.}} Charlie randomly chooses recorded detection times and requires Alice and Bob to alternatively announce their classical bits in the chosen time slots through an authenticated classical channel. According to their correlation of bits, Charlie will get the quantum bit error rate (QBER). Then Charlie will make a decision whether they discard all their bits and restart the whole QSS at Step $1$.

~\noindent{\it{4.~Post-processing.}} Alice, Bob and Charlie conduct classical error correction and privacy amplification to distill the final full key and partial keys.

\begin{figure}[t!]
  \centering
  \includegraphics[width=86mm]{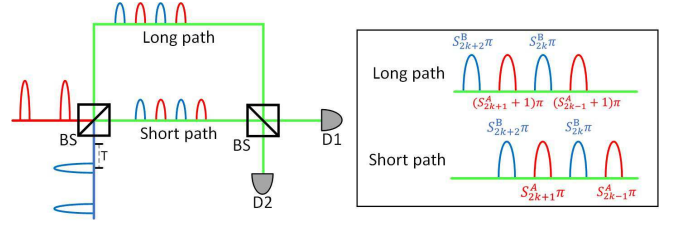}
  \caption{\textbf{Interference measurement at Charlie's site.}} After acted by the former beam splitter, pulse trains from Alice and Bob with the period of 2T will be transformed into pulse trains with the period of T in two paths, respectively. In both paths, pulses at odd time slots are from Alice and pulses at even time slots are from Bob. Phases of pulses from Alice in the long path will be modulated by $\pi$. Then, pulses from two paths will interfere with each other after the delay of time T in the long path. Every pulse from Alice (Bob) will interference with both the former and the later pulses from Bob (Alice).
  \label{charlie}
\end{figure}

As shown in Fig. \ref{charlie}, we demonstrate the bit correlation among Alice, Bob and Charlie. We denote $S_{k}^{X} \in \{0,1\} ~(X\in\{A, B, C\})$ as a classical bit, whose superscript denotes who holds this bit and whose subscript denotes its sequence in the combined pulse train. At Step $1$, when ignoring the global phase of each pulse, we easily find weak coherent states sent by Alice (Bob) can be written as $\bigotimes_{k=1}^{N}|e^{iS_{2k-1}^{A}\pi}\sqrt{\mu}\rangle~\left(\bigotimes_{k=1}^{N}|e^{iS_{2k}^{B}\pi}\sqrt{\mu}\rangle\right)$. Alice (Bob) holds $N$ logic bits denoted by $\{ S_{2k-1}^{A} \}~ \left(\{ S_{2k}^{B} \} \right) $. With an asymmetric interferometer, we can conclude correlation between three participants as:
\begin{equation}
\begin{aligned}
&1 \oplus {S}_{2k-1}^{C} = S_{2k-1}^{A} \oplus S_{2k}^{B},\\
&S_{2k}^{C} = S_{2k}^{B} \oplus S_{2k+1}^{A}.\\
\end{aligned}
\end{equation}

For example, under the circumstance that $j = 2k$, when the modulated phases in one time slot imposed by Alice and Bob are \{0, 0\} or \{$\pi$, $\pi$\}, the differential phase is 0 and D1 clicks. When Alice's and Bob's modulation phases are \{0, $\pi$\} or \{$\pi$, 0\}, the differential phase is $\pi$ and D2 will click. Thus, Charlie's logic bits are the exclusive OR of Alice's and Bob's logic bits. When $j = 2k-1$, to achieve the same correlation, Charlie should bit-flip his corresponding classical bit. Evidently, here Alice and Bob will know Charlie's key bits only by cooperation.

\section{Security Analysis}

The security of our protocol against eavesdropping is discussed in this section. First, we will build an equivalence between our protocol and  differential phase shift QSS in \cite{inoue2008differential}. Then, under the equivalence, both an external eavesdropper and an internal eavesdropper are considered using the conclusion in  differential phase shift quantum key distribution~\cite{waks2006security}.

\subsection{The equivalence}

\begin{figure}
  \centering
  \includegraphics[width=86mm]{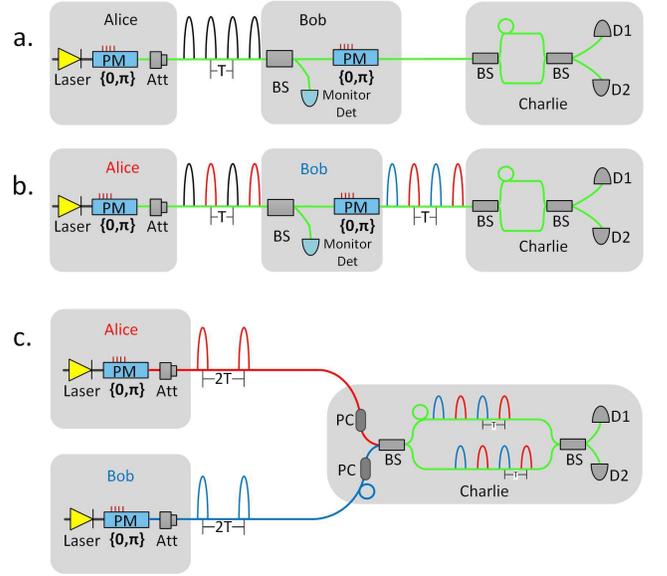}
  \caption{\textbf{Configuration of equivalence between differential phase shift QSS and our protocol.} (a) We present a typical setup of differential phase shift QSS.  (b) With some special rules, differential phase shift QSS will be equivalent to our protocol. (c) For clearly revealing the relationship, we also plot the setup of our protocol.}
  \label{oetoours}
\end{figure}

In Fig.\ref{oetoours} we reduce our scheme to differential phase shift QSS, which is shown in Fig.\ref{oetoours}a. The laser source at Alice's site generates a bunch of weak coherent states, which are all phase-modulated with $\phi_{j}^{A'} \in \{0, \pi\}$, where, without loss of generality, we also set $\mu$ as the intensity of each pulse and the global phase is ignored. Here we also use $j\in \{1,2,3,...\}$ to number the time slots of each pulse. When reaching Bob's site, some of photons will be split to reveal whether Alice is a malicious participant~\cite{inoue2008differential}. The remaining photons will be phase-modulated again by Bob with $\phi_{j}^{B'} \in \{0, \pi\}$. Before transmitted to Charlie, weak coherent states will be $|{e^{i(\phi_{j}^{A'}+\phi_{j}^{B'})}\sqrt{\mu'}}\rangle$. If the differential phase is 0 ($\pi$), Charlie will record the classical bit as 0 (1) and the corresponding detection time slot. We easily derive correlations between phase modulation and Charlie's detection results. For $i \in \{1,2,3,...\}$, $S_{j}^{C'}\pi = |(\phi_{j}^{A'}-\phi_{j+1}^{A'}+(\phi_{j}^{B'}-\phi_{j+1}^{B'})|$, where $S$ also denotes the logic bits. The correlation of classical bits in Fig.~\ref{oetoours}a is $S_{j}^{C'}\pi = S_{j}^{A'} \oplus S_{j+1}^{A'} \oplus S_{j}^{B'} \oplus S_{j+1}^{B'}$.

As an intermediate step, a special rule is added for our equivalence where Alice will only modulate phases of the corresponding pulses if $j=2k-1$ and Bob will only phase-modulate the corresponding pulses if $j=2k$. The configuration of this rule is shown in Fig.\ref{oetoours}b. Then, we will derive the correlation as the following form:
\begin{equation}
\begin{aligned}
&S_{2k-1}^{C'} = S_{2k-1}^{A'} \oplus S_{2k}^{B'},\\
&S_{2k}^{C'} = S_{2k}^{B'}\oplus S_{2k+1}^{A'},\\
\end{aligned}
\end{equation}
which is the same correlation after Charlie bit-flips his logic bits when $j=2k-1$ in our protocol. From Fig.~\ref{oetoours}a to Fig.~\ref{oetoours}b, equivalence can be built since in differential phase shift QSS, we only care about differential phases $\phi_{j}^{A'}-\phi_{j+1}^{A'}~(\phi_{j}^{B'}-\phi_{j+1}^{B'})$ of Alice (Bob) and the whole scheme stays unchanged with the phase of one pulse $\phi_{j+1}^{A'}~(\phi_{j}^{B'})$ fixed. Moreover, since in Fig.~\ref{oetoours}b, pulses at even time slots will not carry any information before reaching Bob's site, it will introduce no differences that Bob sends pulses at even time slots, which is equivalent to our protocol. Then equivalence between Fig.~\ref{oetoours}b to Fig.~\ref{oetoours}c can be built. Our protocol is equivalent to differential phase shift QSS.

Besides, in Fig.\ref{oetoours}a, phases of Alice's pulses will be modulated by Bob, with pulses passing through Bob's site. In ~\cite{gisin2006trojan,jain2014trojan}, we know this kind of schemes is not able to prevent powerful Trojan horse attacks and our protocol will fix the security flaw, obviously.

\subsection{External eavesdropping}

First, we easily derive that beam-splitting attacks and intercept-resend attacks by Eve will fail~\cite{inoue2003differential}. Then, as for a general individual attack, here we can assume that Eve will conduct the same attack in differential phase shift QSS~\cite{inoue2008differential}. Therefore, information leakage to Eve is given by a fraction $2\mu(1-\eta)$ of the sifted key after reconciliation in Step 2, related to the intensity of each pulse $\mu$ and the transmittance $\eta$~\cite{waks2006security}.

\subsection{Internal eavesdropping}

Here we have to be wary of malicious Alice or Bob. Without loss of generality, we let Bob be the malicious participant.

\begin{figure}
  \centering
  \includegraphics[width=86mm]{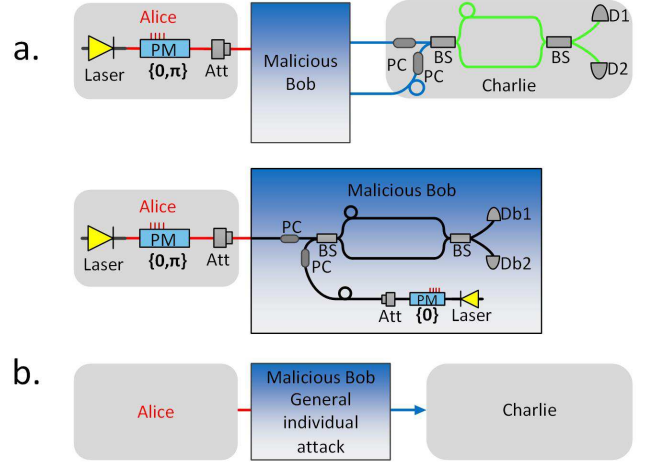}
  \caption{\textbf{Configuration of eavesdropping by malicious Bob.} We describe the attacking strategy of malicious Bob. (a) We present a specific individual attack as discussed in~\cite{inoue2008differential} where Bob will also send a pulse train with 2T period without phase-modulation. (b) A general individual attack by Bob is presented.}
  \label{mB}
\end{figure}

When discussing Bob's eavesdropping, we know he wants to know Charlie's key and also has to know Alice's bits to pass the test-bit checking in Step 4 since Alice and Bob alternatively announce their test bits, introducing a fraction of ${1}/{2}$ of error rates. The configuration for Bob's individual attacks is shown in Fig.~\ref{mB}. He splits part of photons in Alice's signal to be measured after Charlie discloses his detection time. Besides, in the presence of channel noise Bob can potentially attack the rest of pulse trains by a beam-splitting attack. Bob can also entangle every photon transmitted to Charlie with an independent probe to eavesdrop information. To sum up, there are two parts of the eavesdropping strategy which have to be addressed. The first is how much information can be extracted from the split photons and the beam-splitting attacks. The other is information leakage to Bob by his probe states.

For photon-splitting attacks,  malicious Bob only requires to measure differential phases of adjacent pulses $\Delta\phi=\phi_{i}^{A'}-\phi_{i+1}^{A'}$ generated by Alice based on Charlie's detection times. However, without phase reference, malicious Bob in our protocol has to measure the exact phases $\phi_{i}^{A}$ Alice modulates. In that case, we assume malicious Bob will adopt the following attacking strategy in Fig.~\ref{mB}a. To obtain Alice phase-modulation information, Bob applies the same weak laser source to interfere with Alice's pulses, just like what will happen in Charlie's site. Without phase modulation on his eavesdropping pulses, he will get the exact phase information of Alice. Without loss of generality, intensity of Bob's eavesdropping pulses is also $\mu$. Here, we just offer an explicit individual attack as an example.

Then errors induced by split photons is considered as follows. Probability that Bob knows Alice's bit corresponding to Charlie's bit is $2\mu$, and the bit error rate is $(1-2\mu)/{2}$. $\beta$ is the upper bound for the allowable rate of Bob's photon-splitting attacks\cite{waks2006security}. Provided that the total system error rate is $E_{\mu}$, $\beta$ is given by
\begin{equation}\label{eq1}
\beta\left(\frac{1-2\mu}{2}\right)\frac{1}{2}=E_\mu,
\end{equation}
where ${1}/{2}$ means that Alice and Bob disclose their key bits used for test-checking alternatively and then a bit error is revealed when Bob discloses his bit first. For the remaining photons of Alice' signal, Bob conducts a beam-splitting attack utilizing the transmission loss from Alice to Charlie. Provided that the transmittance from Alice to Charlie is $\eta$, Bob stores $(1-\eta)$ of Alice's pulse train and makes his own pulse train without phase modulation interfere with it pulse by pulse after Charlie discloses the photon detection time. This beam-splitting attack gives Bob partial information with a ratio of $2\mu(1-\eta)(1-\beta)$. Then, Bob obtains $2\mu\beta + 2\mu(1-\beta)(1-\eta)$ of Charlie's key in total.

In Fig.~\ref{mB}b, we also consider general individual attacks by Bob. Based on the equivalence between our protocol and differential phase shift QSS, here individual attacks for eavesdropping differential phases stays unchanged in the case where one of two phases is fixed. Then Bob conducts general individual attacks as carried out by Eve in differential phase shift quantum key distribution~\cite{waks2006security}, where the fraction Bob obtains about Charlie's bits is $2\mu$. As demonstrated in~\cite{inoue2008differential}, general individual attacks are more powerful than eavesdropping depicted in Fig.~\ref{mB}a. Here we just consider general individual attacks by malicious Bob.

For probe states of Bob, we will directly obtain total information leakage to Bob in the next section when deriving the final key rate of our protocol.

\section{Final Key Rate}\label{final}
In the above analysis, we obtain that the probability information leakage to Eve and malicious Bob is $2\mu(1-\eta)$ and $2\mu$.

\begin{figure}%[ht!]
  \centering
  \includegraphics[width=86mm]{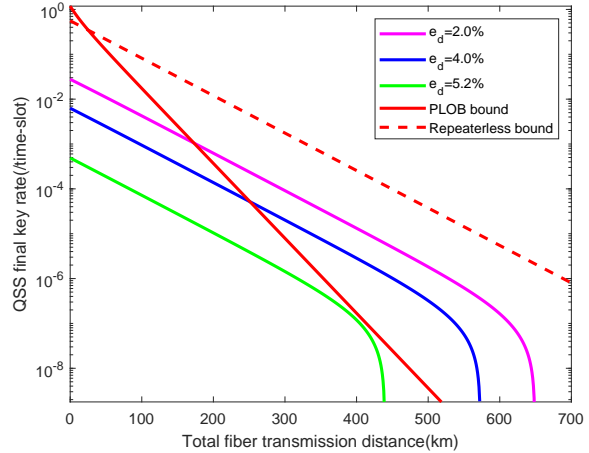}
  \caption{\textbf{Quantum secret sharing key rate vs Transmission distance.} Quantum secret sharing key rate using the experimental parameters. We define the misalignment rate $e_{\rm d}$=2$\%$, $4\%$, 5.2$\%$ to compare their key rates with the PLOB bound between Alice and Bob, together with the repeaterless bound.}
  \label{key}
\end{figure}

We discover that information leakage to external Eve is slightly lower than that to malicious Bob. For simplicity, we can just consider information leakage in our protocol to be $2\mu$. The final key rate of our protocol is
\begin{equation}
R_{\rm QSS} = Q_{\mu}[ -(1-2\mu)\log_2(P_{\rm co}) - f(E_{\mu})h(E_{\mu}) ],
\end{equation}
where $Q_{\mu} = 1-(1-2p_{\rm d})e^{-\mu \eta}$ is the gain of the whole system for Charlie's detections. $h(x)=-x\log_{2}(x)-(1-x)\log_{2}(1-x)$ is Shannon entropy and $f(E_\mu)$ is the error correction efficiency. $P_{\rm co}$ is the upper bound of collision probability when considering individual attacks, which can be concluded as $P_{\rm co} = 1-E_{\mu}^2-{(1-6E_{\mu})^2}/{2}$\cite{lutkenhaus1999estimates}.
For two detectors used by Charlie, we derive total dark count rate as $2p_{\rm d}$ and the error rate of background $e_0 = {1}/{2}$. The total gain and the total error rate with an intensity of $\mu$ is given by
$Q_\mu = 1 - (1-2p_{\rm d})e^{-\mu \eta}$ and
$E_{\mu}Q_{\mu} = e_{\rm d} Q_\mu + \left(\frac{1}{2}-e_{\rm d}\right) 2p_{\rm d} e^{-\mu \eta}$,
where $e_{\rm d}$ is the misalignment error rate of detectors.

Here, we set that the channel transmittance between Alice and Charlie is the same as that between Bob and Charlie. The total distance between Alice and Bob is $L$ to present our transmission rate. In that scheme, transmittance $\eta$ becomes $\eta_{\rm d} \times 10^{-{\alpha L}/{20}}$, where $\eta_{\rm d}$ is detection efficiency of Charlie's detectors. Therefore, distance between Alice and Charlie is ${L}/{2}$ so as to break the linear PLOB bound and the repeaterless bound between Alice and Bob, where formulas of the PLOB bound and the repeaterless bound is $\eta_{\rm PLOB}=-\log_2(1-\eta_{\rm d}\times10^{-{\alpha L}/{10}})$ and $\eta_{\rm repeaterless}=\eta_{\rm d}\times10^{-{\alpha L}/{20}}$. We optimize our transmission rate over the free parameter $\mu$ according to distance and utilize parameters shown in Table.\ref{table01}

\begin{table}%[ht]
\centering
\caption{Simulation parameters. $\eta_{\rm d}$ and $p_{\rm d}$ are the detection efficiency and dark count rate. $\alpha$ is the attenuation coefficient of the ultra-low fiber. $f$ is the error correction efficiency.}\label{table01}
 \setlength{\tabcolsep}{7mm}{
\begin{tabular}{c|c|c|c}
\hline
\hline
$\eta_{\rm d}$ & $p_{\rm d}$ & $\alpha$ & $f$ \\
\hline
$56\%$ & $10^{-8}$ & 0.167 & 1.16 \\
\hline
\hline
\end{tabular}}
\end{table}

\begin{figure}%[ht!]
  \centering
  \includegraphics[width=86mm]{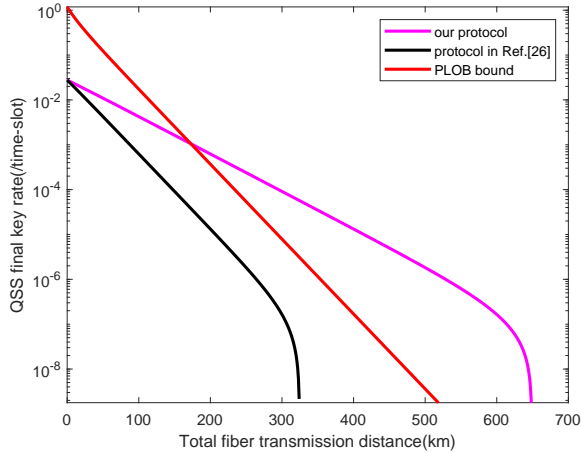}
  \caption{\textbf{Comparison between our protocol and DPS-QSS in Ref.~\cite{inoue2008differential}.}}
  \label{comparison}
\end{figure}

For the intensity $\mu$, we utilize Genetic Algorithm to obtain its optimal value under a certain distance.
We present our simulation result in Fig.\ref{key}. Evidently, when the misalignment rate $e_{\rm d}$ is lower than 5.2\%, our protocol will break the linear PLOB bound\cite{pirandola2017fundamental}. Therefore, theoretical transmission rate of our protocol will reach a level of 600km. For comparison with differential phase shift QSS, we also plot results of our protocol and results in differential phase shift QSS~\cite{inoue2008differential}. With the same $e_{\rm d}$ as $2.0\%$, we can see clearly from Fig.\ref{comparison} that our protocol is much better, increasing the final key rate by about three orders of magnitude in a 300-km-long fiber.

\section{Conclusion}

In summary, we have presented a quantum secret sharing protocol with three participants which is secure against individual attacks. The key rate of our protocol will break the linear PLOB bound, whose theoretical transmission distance will approach 600 km. Compared with the former differential phase shift QSS, our protocol can improve the final key rate by about three orders of magnitude in a 300-km-long fiber and can be secure against Trojan horse attacks. In addition, our protocol requires simple apparatus setup and make experimental and practical implementations more accessible using current techniques.
A full security analysis considering collective attacks and general coherent attacks allowed by quantum mechanics will be the future work. We believe that our protocol will promote the commercialization of quantum secret sharing and will be of wide use to future quantum networks.

\section{Acknowledgments}
We gratefully acknowledge support from the National Natural Science Foundation of China (under Grant No. 61801420); the Key Research and Development Program of Guangdong Province (under Grant No. 2020B0303040001); the Fundamental Research Funds for the Central Universities.

\bibliographystyle{apsrev}

\end{document}